\documentclass{elsart}

\input epsf

\begin{document}

\begin{frontmatter}

\title{Study of the  $D^0 \rightarrow K^+K^-\pi^+\pi^-$ decay.}

\collaboration{The FOCUS Collaboration}

\author[ucd]{J.~M.~Link}
\author[ucd]{P.~M.~Yager}
\author[cbpf]{J.~C.~Anjos}
\author[cbpf]{I.~Bediaga}
\author[cbpf]{C.~G\"obel}
\author[cbpf]{A.~A.~Machado}
\author[cbpf]{J.~Magnin}
\author[cbpf]{A.~Massafferri}
\author[cbpf]{J.~M.~de~Miranda}
\author[cbpf]{I.~M.~Pepe}
% ...

\address[ucd]{University of California, Davis, CA 95616}
\address[cbpf]{Centro Brasileiro de Pesquisas F\'{\i}sicas, Rio de Janeiro, RJ, Brasil}
% ...

\begin{abstract}
Using data from the FOCUS (E831) experiment at Fermilab, we present a new 
measurement for the
Cabibbo-suppressed decay mode $D^0 \rightarrow K^+K^-\pi^+\pi^-$. We measure:

$\Gamma(D^0 \to K^+K^-\pi^+\pi^-)/\Gamma(D^0 \to K^-\pi^-\pi^+\pi^+) = 0.0295 
 \pm 0.0011 \pm 0.0008\,.$
   
 An amplitude analysis has been performed in order to determine the resonant 
substructure of this decay mode. The dominant components are the decays

$D^0 \to K_1(1270)^+ K^-$, $D^0 \to K_1(1400)^+ K^-$ and $D^0 \to \rho(770)^0 \phi(1020)$.

\end{abstract}

\end{frontmatter}  

\textbf{1. Introduction}

In recent years there has been a growing interest in multi-body
hadronic decays of charm mesons. These decays provide a unique tool for investigating the
weak decay of the charm quark in the environment of low energy strong
interactions.
High statistics data with small backgrounds are still dominated by three-body
final states, but results on four-body final states are becoming available.

The picture emerging from many different Dalitz plot analyses
reveals a rich resonant structure of these decays  with a well defined pattern: 
final states that can proceed via simple spectator amplitudes in which the
virtual $W$ 
is coupled to a vector (V) or axial-vector (A) meson have large branching fractions 
compared to cases in which the $W$ is coupled to a pseudoscalar (P) meson. 

The decay mode $D^0 \to K^+K^-\pi^+\pi^-$ is Cabibbo-suppressed and may be produced through
two and three-body intermediate resonant states. Following the above mentioned pattern, 
one expects dominant contributions from modes having the
axial-vector mesons $K_1$ ($D \to AP$), namely $D^0 \to K_1(1270)^+K^-$ and $D^0 \to K_1(1400)^+K^-$, 
both resulting from $W$-radiation (spectator) amplitudes. Other
tree-level amplitudes lead to intermediate two-body decays such as $D \to \phi(1020)\rho(770)^0 $ 
($D \to VV$), or even three-body decays like $D^0 \to \rho(770)^0 K^+K^-$ ($D \to VPP$). The other 
possible contribution of the $D \to VV$ type comes from the mode  $D^0 \to
K^*(892)^0\bar K^*(892)^0$. 
In this case, however, there is no tree-level amplitude and  
the two $W$-exchange amplitudes cancel almost exactly in the SU(3) limit.
This is an instance of a mode that should proceed primarily through final state interactions.  

Most theoretical predictions of hadronic decay rates are still limited to Cabibbo favored 
modes and  two-body decays. There are several predictions for the rate of 
the $D^0 \to \rho(770)^0 \phi(1020)$ mode 
\cite{BSW,BEDAQUE}, but we are unaware of any prediction for $D^0 \to K_1^+K^-$.

 In this paper we present a new measurement of the branching ratio
 $\Gamma(D^0 \to K^+K^-\pi^+\pi^-)/\Gamma(D^0 \to K^-\pi^-\pi^+\pi^+)$ 
using data from the FOCUS experiment.  An amplitude analysis has been 
performed to determine the $D^0 \to K^+K^-\pi^+\pi^-$ resonant substructure. 

FOCUS, an upgraded version of E687~\cite{spectro}, is a charm photoproduction experiment 
 which collected 
data during the 1996--97 fixed target run at Fermilab. Electron and positron beams 
(typically with
$300~\textrm{GeV}$ endpoint energy) obtained from the $800~\textrm{GeV}$ 
Tevatron proton beam produce, by means of bremsstrahlung, a photon beam which
interacts with a segmented BeO target~\cite{photon}. 
The mean photon energy for reconstructed charm events
is $\sim 180~\textrm{GeV}$. A system of three multi-cell threshold \v{C}erenkov
counters performs the charged particle identification, separating kaons from
pions up to a momentum of $60~\textrm{GeV}/c$. Two systems of silicon microvertex
detectors are used to track particles: the first system consists of 4 planes
of microstrips interleaved with the experimental target~\cite{WJohns} and the
second system consists of 12 planes of microstrips located downstream of the
target. These detectors provide high resolution in the transverse plane
(approximately $9~\mu\textrm{m}$), allowing the identification and separation of charm
primary (production) and secondary (decay) vertices. The charged particle
momentum is determined by measuring the deflections in two magnets of
opposite polarity through five stations of multi-wire proportional chambers.

%

\vskip 0.5cm \textbf{2. Analysis of the decay mode $D^0 \to K^+K^-\pi^+\pi^-$} %

The final states are selected using a candidate driven vertex
algorithm~\cite{spectro}. A secondary vertex is formed from the four
candidate tracks. The momentum of the resultant $D^{0}$ candidate is used as
a \textit{seed} track to intersect the other reconstructed tracks and to 
search for a primary vertex. The confidence levels of both vertices are
required to be greater than 1\%. Once the production and decay vertices are 
determined, the distance $L$ between the vertices and its error $\sigma _{L}$ are computed. 
This is the most important variable
for separating charm events from non-charm prompt backgrounds. Signal quality is further 
enhanced by cutting on \emph{Iso2}. This isolation
variable requires that all remaining tracks not assigned to the 
primary and secondary vertex have a confidence level smaller than the cut to form a vertex 
with the $D$ candidate daughters. To minimize systematic errors on the measurements of the 
branching ratio,
we use identical vertex cuts on the signal and normalizing mode, namely
$L$\thinspace /\thinspace $\sigma _{L}$ $>$ $9$ and \emph{Iso2} $<$ 10 \%. We also 
require the  primary vertex 
to be formed with at least two reconstructed tracks in addition to the $D^0$ seed. 

The only difference in the selection criteria between the  $D^0 \to K^+K^-\pi^+\pi^-$ and
$D^{0} \to K^-\pi^-\pi^+\pi^+$ decay modes 
lies in the particle identification cuts. The \v{C}erenkov identification cuts used in
FOCUS are based on likelihood ratios between the various particle
identification hypotheses. These likelihoods are computed for a given track
from the observed firing response (on or off) of all the cells that are
within the track's ($\beta =1$) \v{C}erenkov cone for each of our three 
\v{C}erenkov counters. The product of all firing probabilities for all the cells
within the three \v{C}erenkov cones produces a $\chi ^{2}$-like variable 
$W_{i}=-2\ln (\mathrm{Likelihood})$ where $i$ ranges over the electron, pion,
kaon, and proton hypotheses~\cite{cerenkov}. All kaon tracks are required
to have $\Delta _{K}=W_{\pi }-W_{K}$  greater than $3$ and all 
 pion tracks are required to be separated by less than $5$ units from the best
hypothesis, that is $picon=W_\mathrm{min}-W_{\pi }$.
 
Using the set of selection cuts just described, we obtain the invariant mass distribution 
for $K^{-}K^{+}\pi^{-}\pi^{+}$ shown in Fig.~1a. Although the \v{C}erenkov cuts 
considerably reduce  the reflection peak (from $D^0 \to K^-\pi^+\pi^-\pi^+$) to the right of the
signal peak, there is still a distortion of the background due to this surviving 
contamination. The shape of this reflection peak has been determined by generating Monte Carlo
$D^0 \to K^-\pi^+\pi^-\pi^+$ events and reconstructing them as $K^{-}K^{+}\pi^{-}\pi^{+}$.
The mass plot is fit with a function that includes two Gaussians with the same mean but 
different sigmas to take into account the variation in resolution vs.\ momentum of our 
spectrometer~\cite{spectro}, a second-order polynomial for the combinatorial background and a
shape for the reflection obtained by the Monte Carlo simulation. The amplitude of the reflection
peak is a fit parameter while its shape is fixed. A log-likelihood fit gives a signal of 
$2669 \pm 101$ $K^{-}K^{+}\pi^{-}\pi^{+}$ events.

The large statistics $K^{-}\pi^{-}\pi^{+}\pi^{+}$ mass plot is fit with two Gaussians plus a 
second-order polynomial. The fit gives a signal of $131\,763 \pm 453$  $K^{-}\pi^{-}\pi^{+}\pi^{+}$ events. 

The fitted $D^0$ masses are in good agreement with the world average~\cite{PDG} 
and the resolutions are in good agreement with those of our Monte Carlo simulation. 

%
%

\vskip 0.5cm \textbf{3. Relative Branching Ratio}

The evaluation of relative branching ratios requires yields from the fits to
be corrected for detector acceptance and efficiency. These differ among the various
decay modes because of differences in both spectrometer acceptance (due to
different $Q$ values for the two decay modes) and \v{C}erenkov
identification efficiency.

From the Monte Carlo simulations, we compute the relative efficiencies to 
be: 
${\frac{\epsilon(D^0 \to K^+K^-\pi^+\pi^-) }{\epsilon(D^{0} \to K^-\pi^-\pi^+\pi^+)}}
= 0.688 \pm 0.006$. Using the previous results, we obtain the following 
values for the branching ratio:

$\Gamma(D^0 \to K^+K^-\pi^+\pi^-)/\Gamma(D^0 \to K^-\pi^-\pi^+\pi^+) = 0.0295 \pm 0.0011$.

 Our final measurements have been tested by modifying each of the vertex
and \v{C}erenkov cuts individually. The branching ratio is stable
versus several sets of cuts as shown in Fig.~\ref{Brvscuts}. We varied the confidence
level of the secondary vertex from 1\% to 50\%, \emph{Iso2} from $10^{-6}$ to $1$,
$L$\thinspace /\thinspace $\sigma _{L}$ from $6$ to $20$, $\Delta _{K}$ from $1$ to
$5$ and $picon$ from $-6$ to $-2$ and consider the tight cut set used to study 
the sub-resonant structure (see below).

Systematic uncertainties on branching ratio measurements come from
different sources. We consider four independent contributions to the
systematic uncertainty: the \emph{split sample} component, the \emph{fit
variant} component, the component due to the particular choice of the
vertex and \v{C}erenkov cuts (discussed previously), and the limited statistics 
of the Monte Carlo.

The \emph{split sample} component takes into account the systematics
introduced by a residual difference between data and Monte Carlo, due to a
possible mismatch in the reproduction of the $D^{0}$ momentum and the changing
experimental conditions of the spectrometer during data collection. This 
component has been determined by splitting data
into four independent subsamples, according to the $D^{0}$ momentum range
(high and low momentum) and the configuration of the vertex detector,
that is, before and after the insertion of an upstream silicon system. A technique,
employed in FOCUS and modeled after the 
\emph{S-factor method} from the Particle Data Group~\cite{PDG}, was used 
to try to separate true systematic variations from statistical 
fluctuations. The branching ratio is evaluated for each of the $4~(=2^{2})$ 
statistically independent subsamples and a \emph{scaled error} $\tilde{\sigma}$ (that 
is the errors are boosted when $\chi ^{2}/(N-1)>1$) is calculated.  
The \emph{split sample} error $\sigma_\mathrm{split}$ is defined as the 
difference between the reported statistical error and the scaled error, 
if the scaled error exceeds the statistical error~\cite{brkkpipi}.

Another possible source of systematic uncertainty is the \emph{fit variant}.
This component is computed by varying, in a reasonable manner, the fitting
conditions for the whole data set. In our study we fixed the widths of the 
Gaussians to the values obtained by the Monte Carlo simulation, we changed 
the background parametrization (varying the degree of the polynomial), we removed the
reflection peak from the fit function, and we use one Gaussian instead of two. 
Finally the variation of the computed efficiencies,  both for 
$D^0 \to K^{-}K^{+}\pi^{+}\pi^{-}$ and the normalizing decay mode, due to the different 
resonant substructure simulated in the Monte Carlo has been taken into account. 
The BR values obtained by these variants are all {\em a priori} equally likely, therefore this 
uncertainty can be estimated by the {\it r.m.s.}\ of the measurements~\cite{brkkpipi}.

Analogously to the \emph{fit variant}, the cut component is estimated using 
the standard deviation of the several sets of cuts shown in Fig.~\ref{Brvscuts}.
Actually this is an overestimate of the cut component because the statistics of 
the cut samples are different. 

 Finally, there is a further contribution due to the limited statistics of 
the Monte Carlo simulation used to determine the efficiencies. Adding in
quadrature the four components, we get the final systematic 
errors which are summarized in Table~\ref{err_sist}. 

\begin{table}
\begin{center}
\begin{tabular}{|l|c|}
\hline
{Source}        & {Systematic error}      \\
\hline
{Split sample}  &  0.0\%      \\
{Fit Variant}   &  2.1\%      \\
{Set of cuts}   &  1.7\%      \\
{MC statistics} &  0.9\%      \\ \hline
{Total systematic error} &  2.8\%  \\ \hline
\end{tabular}
\caption{Contribution in percent to the systematic uncertainties of 
the branching ratio $\Gamma(D^0 \to K^-K^+\pi^-\pi^+)/\Gamma(D^0 \to K^-\pi^-\pi^+\pi^+)$.}
\label{err_sist}
\end{center}
\end{table}

 The final result is shown in Table~\ref{comparison} along with a
comparison with the previous determinations.

\begin{table}[h!]
\begin{center}
\begin{tabular}{|l|l|l|}
\hline
Experiment& $\frac{\Gamma(D^0 \to K^+K^-\pi^+\pi^-)}{\Gamma(D^0 \to K^-\pi^-\pi^+\pi^+)}$ &
Events \\ \hline
FOCUS (this result)& $0.0295  \pm 0.0011 \pm 0.0008 $ & $ 2669 \pm 101 $    \\ 
E791~\cite{E791}   & $0.0313  \pm 0.0037 \pm 0.0036 $ & $  136 \pm 15 $     \\
E687~\cite{E687}   & $0.035   \pm 0.004  \pm 0.002  $ & $  244 \pm 26 $     \\ 
ARGUS~\cite{ARGUS} & $0.041   \pm 0.007  \pm 0.005  $ & $  114 \pm 20 $     \\
CLEO~\cite{CLEO}   & $0.0314  \pm 0.010  \pm 0.005  $ & $   89 \pm 29 $     \\ \hline
\end{tabular}
\caption{Comparison with other experiments.}
\label{comparison}
\end{center}
\end{table}

%
%

\vskip 1.5cm \textbf{4. Amplitude analysis of $D^0 \to K^+K^-\pi^+\pi^-$}

A fully coherent amplitude analysis was performed in order to determine the
resonant substructure of the $D^0 \to K^+K^-\pi^+\pi^-$ decay. Previous results
on this mode were obtained from small samples. E687~\cite{E687} did an incoherent analysis,
while the analysis of E791~\cite{E791} quoted only inclusive fractions.

Tighter cuts have been applied in the selection of events for the amplitude analysis. 
We require: $L$\thinspace /\thinspace $\sigma _{L} > 10$, secondary CL $> 0.05$, 
$\Delta _{K}$ $>$ $4$, \emph{Iso2} $<$ $10^{-5}$. In addition the secondary vertex
must
lie outside of the segmented targets to reduce contamination due to  secondary 
interactions. Using this 
set of cuts we obtain the invariant mass distribution shown in Fig.~1b. The mass plot is 
fit with a function that includes two Gaussians, a second-order polynomial for the 
combinatorial
background, and a shape for the reflection peak (similar to the fit of Fig.~1a).
A log-likelihood fit finds $1279 \pm 48$ $K^{-}K^{+}\pi^{-}\pi^{+}$ signal events.

Figure 3 shows the $K^-K^+$, $K^-\pi^+$, and $\pi^-\pi^+$ projections. 
We observe clear signals of the $\phi(1020)$, $\rho(770)^0$ and $K^*(892)^0$. These signals could come
from decays of the axial-vector mesons $K_1(1270)^+$ and $K_1(1400)^+$, or from decays of the type
$D^0 \to VV$ and $D^0 \to VPP$ ($V=\phi(1020)$, $\rho(770)^0$, $K^*(892)^0$, $P=\pi^{\pm},K^{\pm}$).
The $K_1(1270)^+$ decay modes that lead to a $K^{-}K^{+}\pi^{-}\pi^{+}$ final state are  
$\rho(770)^0 K^+$,
$K^*(892)^0 \pi^+$, $K^*_0(1430) \pi^+$ and $\omega K^+$, whereas the
$K_1(1400)^+$ decays only to $K^*(892)^0 \pi^+$. The $\omega K^+$ mode is not considered because
the branching fraction of $\omega \to \pi^+ \pi^-$ is less than 2\%.
These components are distinguishable by the characteristic angular distributions of the 
final state particles. 

In our model we consider two $D \to AP$ modes, $K_1(1270)^+K^-$ and $K_1(1400)^+K^-$.
We also consider contributions from two $D \to VV$ modes, $\phi(1020) \rho(770)^0$
and $K^*(892)^0\bar K^*(892)^0$, and from three $D \to VPP$ modes, $\rho(770)^0 K^+K^-$, 
$\phi(1020) \pi \pi$, and 
$K^*(892)^0 K^+\pi^-$. Finally, we include the mode $f_0(980) \pi^+ \pi^-$ ($D \to SPP$), since
the $f_0(980)$ has a strong coupling to the $K^+K^-$ channel. The mode $f_0(980) K^+K^-$
is not considered because there is no phase space for it. In principle the mode $a_0(980) \pi^+\pi^-$ 
could also contribute, but with the present statistics one cannot distinguish between the $a_0(980)$ 
and $f_0(980)$. Moreover, the dominant component of the $a_0(980)$ is the $\eta \pi$ channel.
A good description of our data is obtained without the non-resonant channel.

The formalism used in this amplitude analysis is a straightforward extension
to four-body decays of the usual Dalitz plot fit technique.
The overall signal amplitude is a coherent sum of the ten individual amplitudes,  
${\mathcal A} = \sum_k c_k A_k$. The amplitudes $ A_k$ are constructed as a product of form
factors, relativistic Breit-Wigner functions, and spin amplitudes which account for angular momentum
conservation.
We use the Blatt-Weisskopf damping factors \cite{blatt}, $F_l$, as form factors ($l$ is
the orbital angular momentum of the decay vertex).
For the spin amplitudes we use the Lorentz invariant amplitudes \cite{mkiii},
which depend both on the spin of the resonance(s) and the orbital angular momentum. The
relativistic Breit-Wigner is

\[
BW = {1 \over {m^2 - m_0^2 + im_0\Gamma(m)}},
\]
where

\[
\Gamma(m) = \Gamma_0 
\frac{m_0}{m}\left(\frac{p^*}{p^*_0}\right)^{2l+1}\frac{F_l^2}{F_{l0}^2}.
\]

In the above equations, $m$ is the  two-body invariant mass, $m_0$ is the resonance 
nominal mass, and $p^*=p^*(m)$ is the breakup momentum at resonance mass $m$.

Masses and widths of resonances are taken from the PDG~\cite{PDG}, except for the  
$\rho(770)^0$ and the $f_0(980)$. The line shape of the $\rho(770)^0$ is taken from
Crystal Barrel~\cite{cbarrel}, and includes the $\rho - \omega$ interference. This significantly
improves the fit.
For the $f_0(980)$ we used the Flatte formula~\cite{flatte} of a coupled channel Breit-Wigner 
function. The $f_0(980)$ parameters --- $g_\pi = 0.20 \pm 0.04$, 
$g_K = 0.50 \pm 0.20$ and $m_0 = (0.957 \pm 0.008) \mathrm{GeV}/c^2$ --- are obtained 
by a fit to the FOCUS 
$D_s^+ \to \pi^+ \pi^- \pi^+$ Dalitz plot, where the $f_0(980) \pi^+$ is the dominant component. 
Magnitudes and phases of the $K_1(1270)^+$ decay modes relative to the $\rho(770)^0 K^+$ were 
determined by the fit, since the limited $D^0$ phase space will affect 
the $K_1(1270)^+$ decay fractions.

The fit parameters are the 9 complex coefficients $c_k$. Magnitudes and phases are relative to 
those of the chain $D^0 \to K_1(1270)^+K^-$, $K_1(1270)^+ \to \rho(770)^0 K^+$.
The overall signal amplitude is corrected on an event-by-event basis 
for the acceptance, which is nearly constant across the phase space.
The finite detector resolution causes a smearing of the edges of the five-dimensional
phase space. This effect is taken into account by multiplying the overall signal distribution 
by a Gaussian factor, $g(M)$ ($M$ being the $K^+K^-\pi^+\pi^-$ mass). The normalized
signal probability distribution is, thus,

\[
P_S(\phi) = \frac {1}{N_S} \varepsilon(\phi) \rho(\phi)
g(M) \left| \sum c_k A_k(\phi) \right|^2 
\]
with $\phi$ being the coordinates of an event in the five dimensional phase space, 
$\varepsilon(\phi)$ the acceptance function, and $\rho(\phi)$ the phase space density.

We consider four
types of background events: random combinations of a $\phi(1020)$ and a
$\pi^-\pi^+$ pair, a $K^*(892)^0$ plus a $K^-\pi^+$ pair, a $\rho(770)^0$ plus a $K^+K^-$ pair,
and random combinations of $K^+K^-\pi^+\pi^-$. The relative fractions of these backgrounds
were determined from a fit to the data on the side bands
of the $K^+K^-\pi^+\pi^-$ mass spectrum. This fit yields 69\% for  random combinations of 
$K^+K^-\pi^+\pi^-$. The fractions of the $\phi(1020)\pi^+\pi^-$, $K^*(892)^0K^-\pi^+$, 
and $\rho(770)^0K^+K^-$  backgrounds are 17\%, 5\% and 9\%, respectively.

We assume the random  
$K^+K^-\pi^+\pi^-$ combinations to be uniformly distributed in phase space, whereas
for the other backgrounds we assume  Breit-Wigners with no
form factors and no angular distribution.
The overall background distribution is a weighted, incoherent 
sum of the four components described above. The relative background fractions, 
$b_k$, are fixed in the fit. The overall background distribution
is also corrected for the acceptance (assumed to be the same as for the signal events) 
on an event-by-event basis and multiplied by an exponential function $b(M)$,  
accounting for the $K^+K^-\pi^+\pi^-$mass distribution of the background. The 
normalized background probability distribution is
 
 \[
P_B(\phi) = \frac{1}{N_B}\varepsilon(\phi) \rho(\phi) b(M) \sum b_k B_k(\phi).
\]

An unbinned maximum likelihood fit was performed, minimizing the quantity
$w\equiv -2\ln(\mathcal{L})$. The likelihood function, $\mathcal{L}$, is

\[
\mathcal{L} = \prod_\mathrm{events} \left[P_S(\phi^i) + P_B(\phi^i)\right].
\]

Decay fractions are obtained from the coefficients $c_k$, determined by the fit,
and after integrating the overall signal amplitude over the phase space~\cite{kkkp}. Errors
on the fractions include errors on both magnitudes and phases, and are computed using the 
full covariance matrix.

The result from the best fit is shown in Table~\ref{fit}. 
The dominant component is indeed the mode $D^0 \to K_1^+(1270^+) K^-$, with a fraction of 
(33 $\pm$ 6 $\pm$ 4)\%. 
Integrating the amplitude squared for the summed $D \to AP$ components
gives nearly 55\% of the total decay rate.
This result is consistent with other $D$ decays, for instance, the case of 
$D^0 \to K^- \pi^+ \pi^- \pi^+$, which is dominated by the mode $a_1(1260)^+ K^-$ with a 
fraction of about 50\%.

The second dominant component is the $D \to VV$ mode:
a large contribution of the $\phi(1020)\rho(770)^0$ channel, with a fraction of 29\%, and a small,
but significant, contribution from $K^*(892)^0 \bar K^*(892)^0$. The $\phi(1020)\rho(770)^0$ channel can
proceed through an internal $W$ radiation amplitude, while the $K^*(892)^0 \bar K^*(892)^0$
results from final state interactions. Using the value of
$\Gamma(D^0 \to K^+K^-\pi^+\pi^-)/\Gamma(D^0 \to K^-\pi^-\pi^+\pi^+)$ we measured,
we obtain $B(D^0 \to \phi(1020) \rho(770)^0) = (1.2 \pm 0.1) \times 10^{-3} $. This is
a factor of five higher than the prediction of Bedaque, Das and Mathur~\cite{BEDAQUE},
($B(D^0 \to \phi(1020) \rho(770)^0) = 2.2 \times 10^{-4}$). The model of Bauer, Steck and 
Wirbel~\cite{BSW}
predicts $B(D^0 \to \phi(1020) \rho(770)^0) = 4.5 \times 10^{-3}$,
which is a factor of 3.5 higher than our value.

Finally, the remaining fraction comes from the $D \to VPP$ and $D \to SPP$ decays. Altogether,
these modes account for nearly 30\% of the total decay rate. A similar result was
found in our analysis
of the $D^0 \to K^+ K^- K^- \pi^+$ decay~\cite{kkkp}, showing  the importance of 
the three-body channels in four-body decays of $D$ mesons. In addition to the three $D \to VPP$
modes listed above, we have also included the mode $K^*(1400)^+K^-$, but its contribution
is negligible.

The  $K^+K^-$ spectrum has an interesting feature.
Near the threshold it is dominated by the $\phi(1020)$ peak, but
the $\phi(1020)$ line shape is distorted by the presence of the $f_0(980)$. 
The $\phi(1020)$ and $f_0(980)$ cannot be distinguished with purely a mass cut on $K^+K^-$ 
invariant mass. The fairly large fraction (15\%) of the $D^0 \to f_0(980) \pi^+ \pi^-$ mode 
shows that the $f_0(980)$ contribution cannot be neglected. The distinction between 
$\phi(1020)$ and $f_0(980)$ components requires a full angular analysis.

In four body decays the phase space is 5-dimensional, so we can only look at projections.
In Figure~\ref{proj},  the $K^+K^-$, $\pi^+ \pi^-$, and  $K^-\pi^+$ invariant mass
projections of events used in the amplitude analysis are superimposed on the fit result, 
with the background projections shown in the shaded histograms. The fit result can also 
be displayed in two dimensional projections, shown in Figure~\ref{proj2}, and in
distributions of the cosine of helicity and acoplanarity angles, shown
in Figure~\ref{ang}. In the 
case of the $D^0 \to \phi(1020) \rho(770)^0$ mode, the helicity angle is defined as the angle 
between the $K^-$ ($\pi^-$) and the direction of the recoiling $\rho(770)^0$ ($\phi(1020)$) in the 
$\phi(1020)$ ($\rho(770)^0$) rest frame. In the case of $D^0 \to  K^*(892)^0 \bar K^*(892)^0$, 
helicity angles are defined as the angle between the $K$ and the recoiling  $K^*(892)^0$ in 
each $K^*(892)^0$ frame.
The acoplanarity angle is the angle between the $\phi(1020)$ and $\rho(770)^0$ decay 
planes, measured in the $D$ rest frame. The bump in the central part of the $\rho(770)^0$ helicity 
angle distribution is due to the decay chain $D^0 \to K_1(1270)^+ K^-$, 
$K_1(1270)^+ \to \rho(770)^0 K^+$.


\begin{table}
\begin{center}
\begin{tabular}{|c|c|c|c|}     \hline
          Mode                   & Magnitude              &   Phase      &   Fraction (\%)
 \\ \hline \hline
 {$K_1(1270)^+K^-, ~K_1 \to \rho(770)^0 K^+$} &  1 (fixed)             &   0 (fixed)               & 18 $\pm$ 6 $\pm$ 3
 \\ \hline  
 {$K_1(1270)^+K^-, ~K_1 \to K^*_0(1430)\pi^+$} & 0.27 $\pm$ 0.08 $\pm$ 0.06 & 354 $\pm$ 19 $\pm$ 19 & 2 $\pm$ 1 $\pm$ 0   
 \\ \hline
 {$K_1(1270)^+K^-, ~K_1 \to K^*(892)^0\pi^+$}  & 0.94 $\pm$ 0.16 $\pm$ 0.13 &  12 $\pm$ 12 $\pm$ 15 & 16 $\pm$ 4 $\pm$ 5   
 \\ \hline
 {$K_1(1270)^+K^-$, (all modes)}    & ---                          &  ---                    & 33 $\pm$ 6 $\pm$ 4
  
 \\ \hline  
 {$K_1(1400)^+K^- $}                 & 1.18 $\pm$ 0.19 $\pm$ 0.09 & 259 $\pm$ 11 $\pm$ 13 & 22 $\pm$ 3 $\pm$ 4  
 \\ \hline
 {$K^*(892)0 \bar K^*(892)^0$}      & 0.39 $\pm$ 0.09 $\pm$ 0.11 &  28 $\pm$ 13 $\pm$ 10 &  3 $\pm$ 2 $\pm$ 1  
 \\ \hline 
 {$\phi(1020)\rho(770)^0$}          & 1.30 $\pm$ 0.11 $\pm$ 0.07 &  49 $\pm$ 11 $\pm$ 12 & 29 $\pm$ 2 $\pm$ 1
 \\ \hline
 {$\rho(770)^0 K^+ K^- $}           & 0.33 $\pm$ 0.12 $\pm$ 0.16 & 278 $\pm$ 26 $\pm$ 20 &  2 $\pm$ 2 $\pm$ 2   
 \\ \hline
 {$\phi(1020) \pi^+ \pi^- $}        & 0.30 $\pm$ 0.06 $\pm$ 0.06 & 163 $\pm$ 16 $\pm$ 15 &  1 $\pm$ 1 $\pm$ 0 
 \\ \hline 
 {$K^*(892)^0 K^+ \pi^-$}           & 0.83 $\pm$ 0.09 $\pm$ 0.10 & 234 $\pm$ 10 $\pm$ 11 & 11 $\pm$ 2 $\pm$ 1    
 \\ \hline
 {$f_0(980) \pi^+ \pi^-$}           & 0.91 $\pm$ 0.13 $\pm$ 0.05 & 240 $\pm$ 11 $\pm$ 17 & 15 $\pm$ 3 $\pm$ 2   
 \\ \hline 
\end{tabular}

\protect\caption {Results from the best fit. The second error on the fractions, magnitudes  
and phases is systematic. The fraction for the mode $K_1(1270)^+K^-$ (fourth row) includes
all three decay modes of the $K_1(1270)^+$, added coherently.}
\label{fit}  
\end{center}
\end{table}  


The invariants used to define the kinematics of the 
$D^0 \rightarrow K^+K^-\pi^+\pi^-$ decay are the  $K^-K^+$, $\pi^+\pi^-$, 
$K^-\pi^+$, $K^+\pi^-$ and $K^-\pi^-$ masses squared. 
 Due to the limited statistics we have integrated
over the latter invariant and divided the remaining four into three bins each, 
yielding a total of 81 cells. From these a total 
of 45 cells lie within the phase space of this decay. 
Data and Monte Carlo samples were divided into these 45 cells. 
The fit has 18 free
parameters, so the number of degrees of freedom is $45-18=27$.
We obtain a $\chi^2$ of 40.4, and from this value the 
estimated confidence level of the fit was 4.7\%.

We consider three different sources of systematic uncertainties for the amplitudes analysis:
\emph{split sample}, using  the same subsamples described previously, \emph{fit variant},
varying the fitting conditions of the whole data set and
the component due to the particular choice of the vertex and \v{C}erenkov cuts.  
The most important contributions from \emph{fit variant} systematic errors are: 
parameterization of the $f_0(980)$ line shape, uncertainty in the relative background
fractions, and uncertainty in the masses and widths of the resonances. We have also considered the effect of
ignoring form factors and using a flat acceptance. The cut component systematic errors
were estimated using the standard deviation of several different sets of vertex/particle 
identification cuts, as we did in the branching ratio measurement. Systematic errors on phases 
and fractions were obtained adding in quadrature the 
\emph{split sample}, \emph{fit variant}, and cut component errors.

%
%

\vskip 1.5cm \textbf{5. Conclusions}

 Using data from the FOCUS (E831) experiment at Fermilab, we studied the Cabibbo-suppressed 
decay mode $D^0 \rightarrow K^+K^-\pi^+\pi^-$.

 A comparison with the two previous determinations of the relative branching ratio 
$\Gamma(D^0 \to K^+K^-\pi^+\pi^-)/\Gamma(D^0 \to K^-\pi^-\pi^+\pi^+)$ shows an 
impressive improvement in the accuracy of this measurement. 

A coherent amplitude analysis of $K^+K^-\pi^+\pi^-$ final states was performed, showing 
that the dominant contribution comes from $D^0 \to AP$ modes, 
corresponding to nearly 55\% of the $D^0 \to K^+K^-\pi^+\pi^-$ decay rate.
We also measured an important contribution from $D^0$ decaying to two vector mesons,
corresponding to nearly 30\% of the decay rate. 
The remaining fraction of the $D^0 \to K^+K^-\pi^+\pi^-$ decay rate
comes from decays of the type $D^0 \to VPP$ and $D^0 \to SPP$, with important contributions 
from $K^*(892)^0  K^- \pi^+ $ and $f_0(980) \pi^+ \pi^-$.

%
%

\vspace{1.cm}

We wish to acknowledge the assistance of the staffs of Fermi National
Accelerator Laboratory, the INFN of Italy, and the physics departments of
the collaborating institutions. This research was supported in part by the
U.~S. National Science Foundation, the U.~S. Department of Energy, the
Italian Istituto Nazionale di Fisica Nucleare and Ministero della Istruzione
Universit\`a e Ricerca, the Brazilian Conselho Nacional de Desenvolvimento
Cient\'{\i}fico e Tecnol\'ogico, CONACyT-M\'exico, and the Korea Research
Foundation of the Korean Ministry of Education.

%
%

%
%

\newpage

\begin{figure}[!!t]
\epsfysize=21.cm \epsfxsize=10.cm \epsfbox{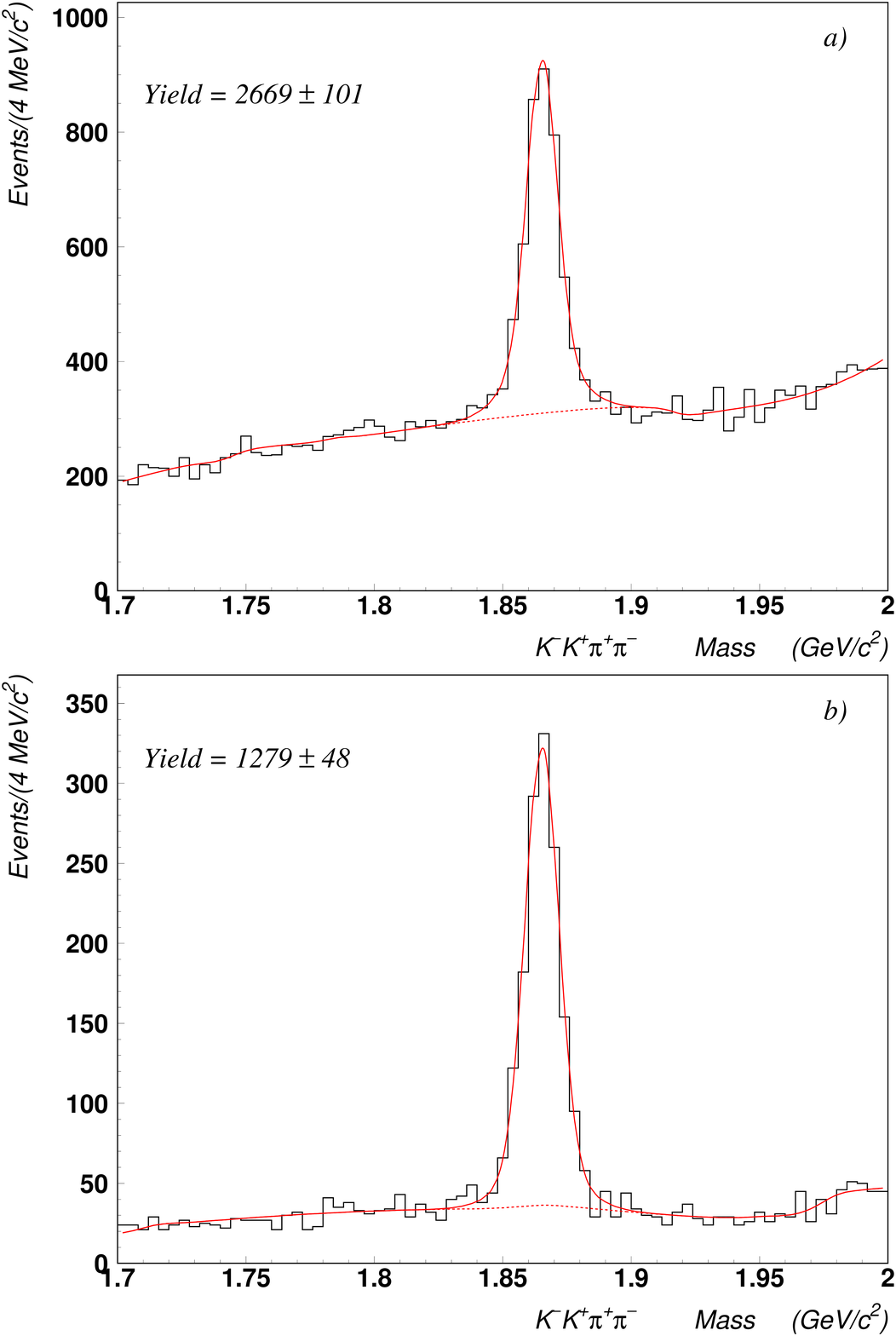} \vspace{0.5cm}
\caption{Invariant mass distributions for $K^-K^+\protect\pi^-\protect\pi^+$
with standard cuts (a) used to determine the BR and tight cuts (b) used 
to study the
subresonant structure. The fit (solid curve) is explained in the text, 
the dashed line
shows the background}
\label{masses}
\end{figure}
\newpage

\begin{figure}[!!t]
\epsfysize=18.cm \epsfxsize=12.5cm \epsfbox{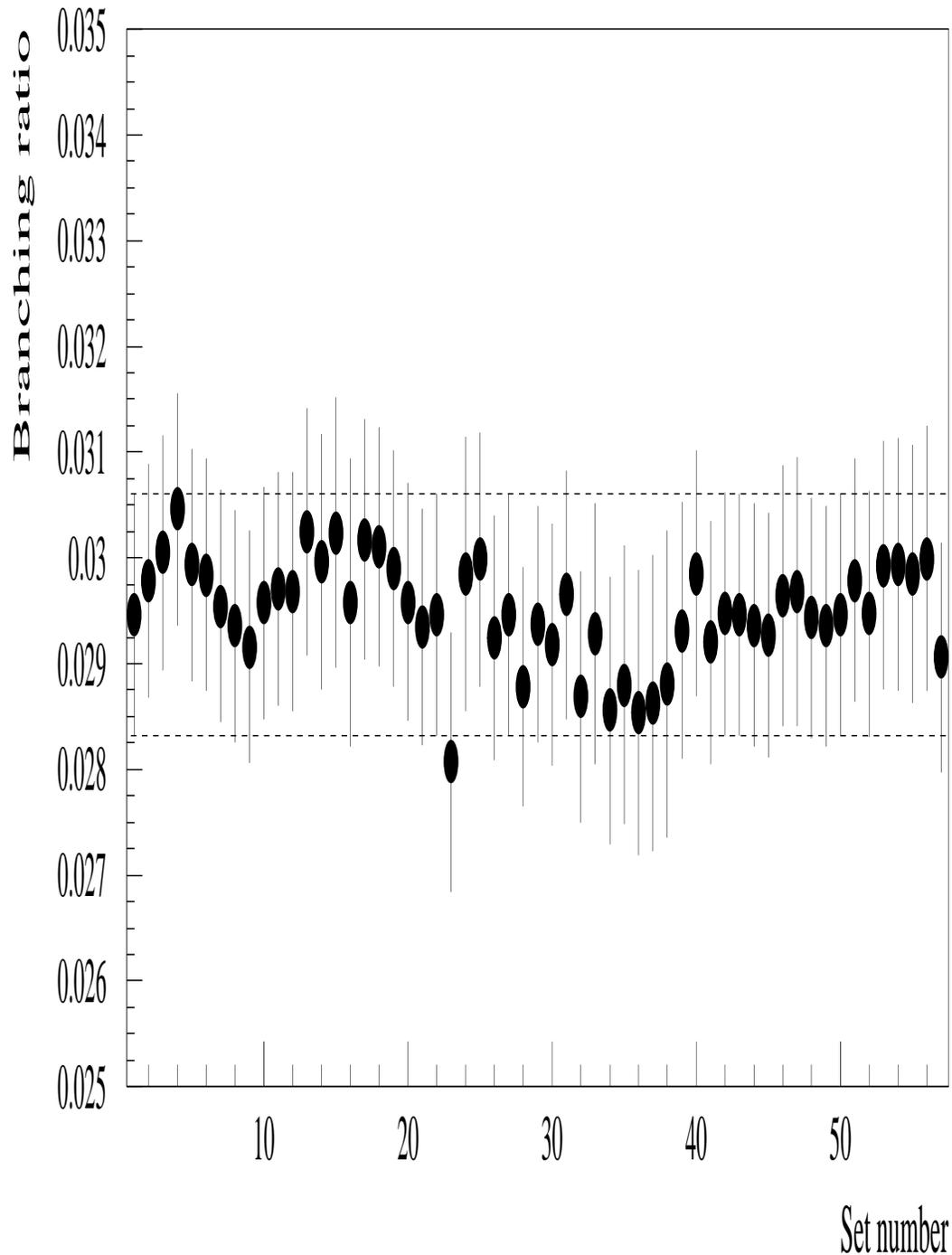} \vspace{0.5cm}
\caption{Branching ratio
$\Gamma(D^0 \to K^-K^+\protect\pi^-\protect\pi^+)/\Gamma(D^0 \to 
K^-\pi^-\pi^+\pi^+)$
versus several sets of cuts. We varied the confidence level of the 
secondary vertex from
$1\%$ to $50\%$ (16 points), \emph{Iso2} from $10^{-6}$ to $1$ (7 points),
$L$\thinspace /\thinspace $\sigma _{L}$ from $6$ to $20$ (15 points),
$\Delta _{K}$ from $1$ to $5$ (9 points), {\it picon} from $-6$ to $-2$ 
(9 points),
and finally we consider the tight cut set used to study the subresonant
structure (last point). The dashed lines show the quoted branching ratio
$\pm 1 \sigma$.}
\label{Brvscuts}
\end{figure}
\newpage

\begin{figure}[!!t]
\epsfysize=18.cm \epsfxsize=12.5cm \epsfbox{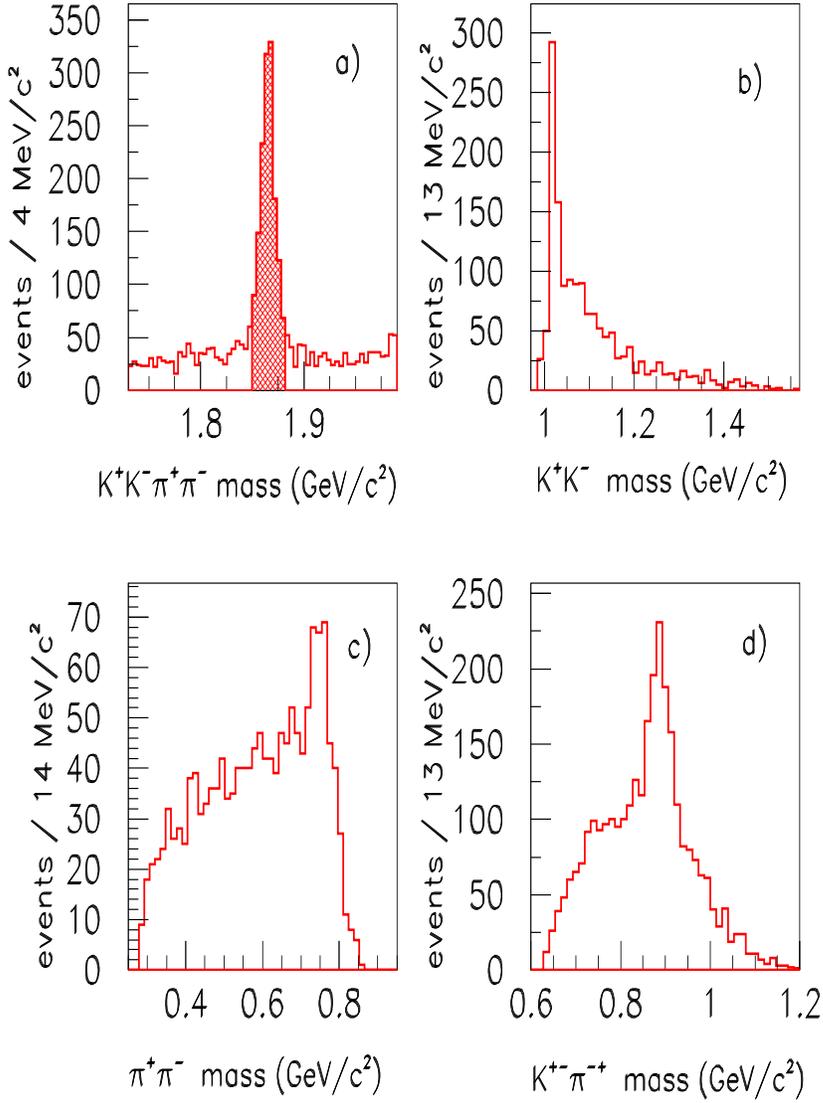} \vspace{0.5cm}
\caption{a) The $K^+ K^-\pi^-\pi^+$ mass plot with tight cuts. 
The hatched area corresponds to the events used in the amplitude analysis.
The other plots show two-body projections of these events: b) $K^+ K^-$; c) $\pi^-\pi^+$;
d) $K^+\pi^-$ plus $K^-\pi^+$ (two independent entries per event).
We observe clear signals of $\phi(1020)$, $\rho(770)^0$ and $K^{*0}(892)$.}
\label{res}
\end{figure}
\newpage

\begin{figure}[!!t]
\epsfysize=18.cm \epsfxsize=12.5cm \epsfbox{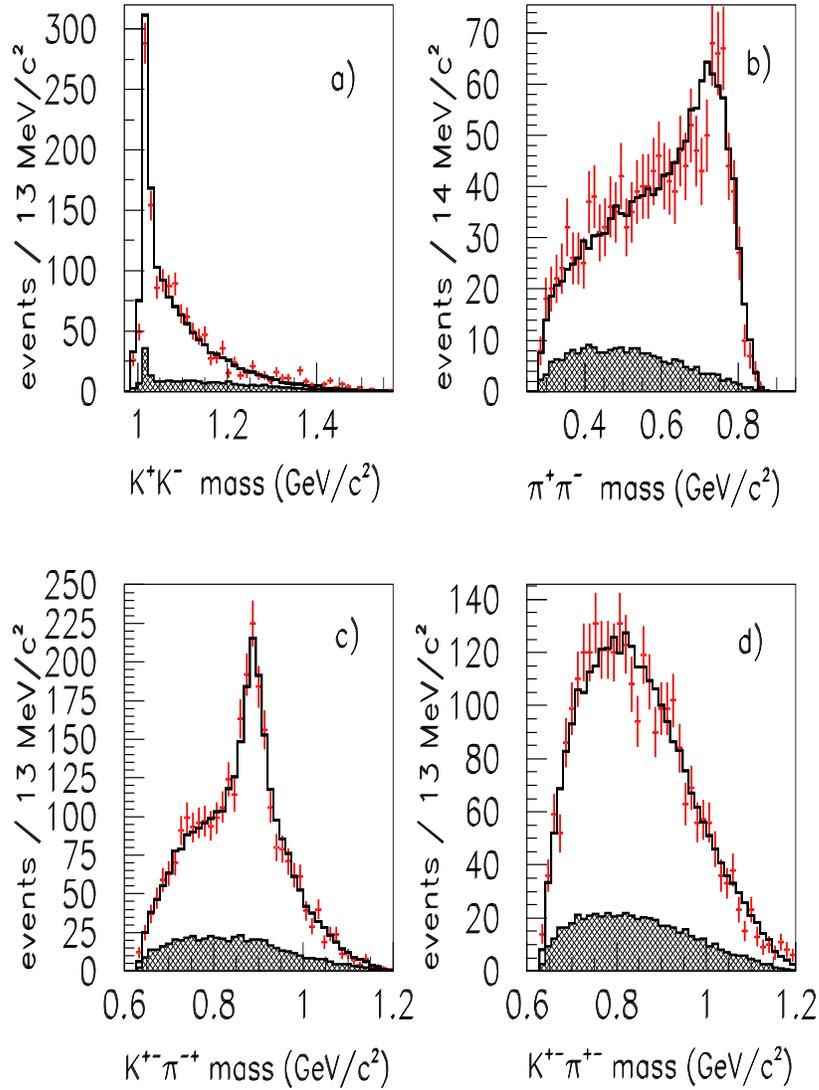} \vspace{0.5cm}
\caption{The two-body projections with the fit result (solid line) and background
(shaded histogram) superimposed: a) $K^+ K^-$; b) $\pi^-\pi^+$; c)
$K^+\pi^-$ plus $K^-\pi^+$ (two entries per event); d)
$K^+\pi^+$ plus $K^-\pi^-$  (two entries per event).}
\label{proj}
\end{figure}
\newpage

\begin{figure}[!!t]
\epsfysize=18.cm \epsfxsize=12.5cm \epsfbox{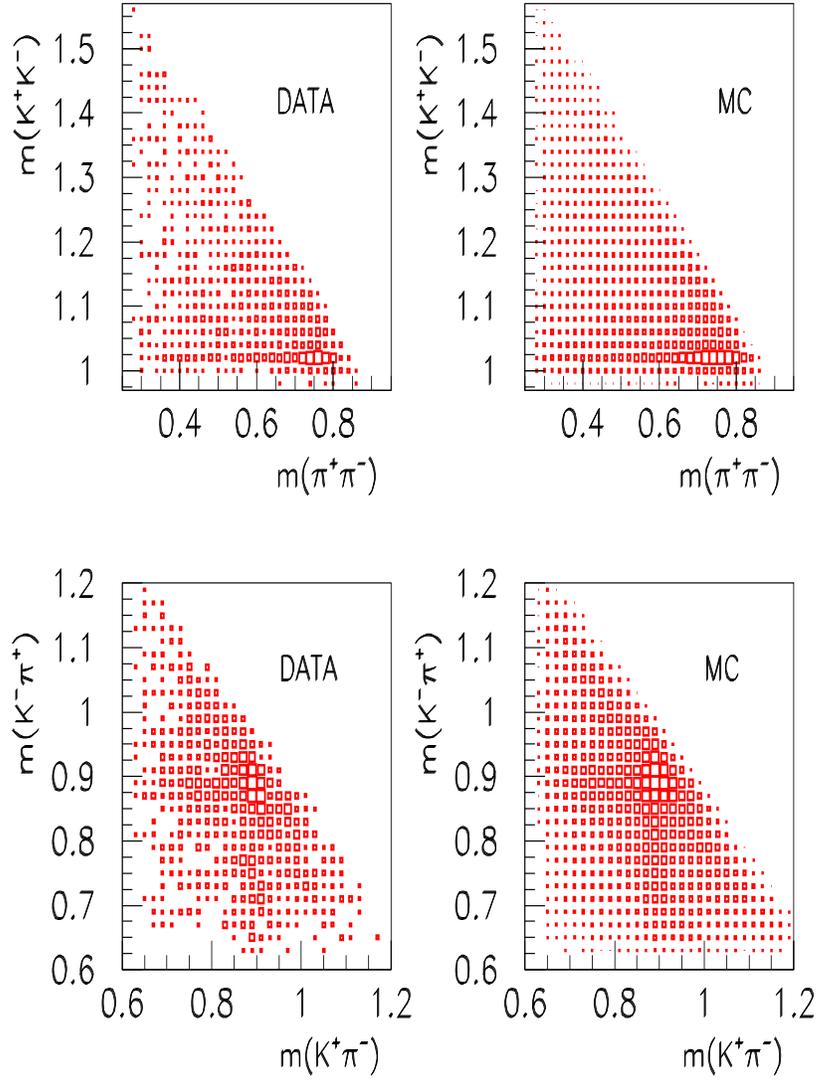} \vspace{0.5cm}
\caption{Two-dimensional projections: $K^+ K^-$ {\em vs.} $\pi^-\pi^+$ in the top row,
$K^+\pi^-$ {\em vs.} $K^-\pi^+$ in the lower row.
Data is in the left column; mini-MC, generated according
to the fit solution, is in the right column. }
\label{proj2}
\end{figure}
\newpage

\begin{figure}[!!t]
\epsfysize=18.cm \epsfxsize=12.5cm \epsfbox{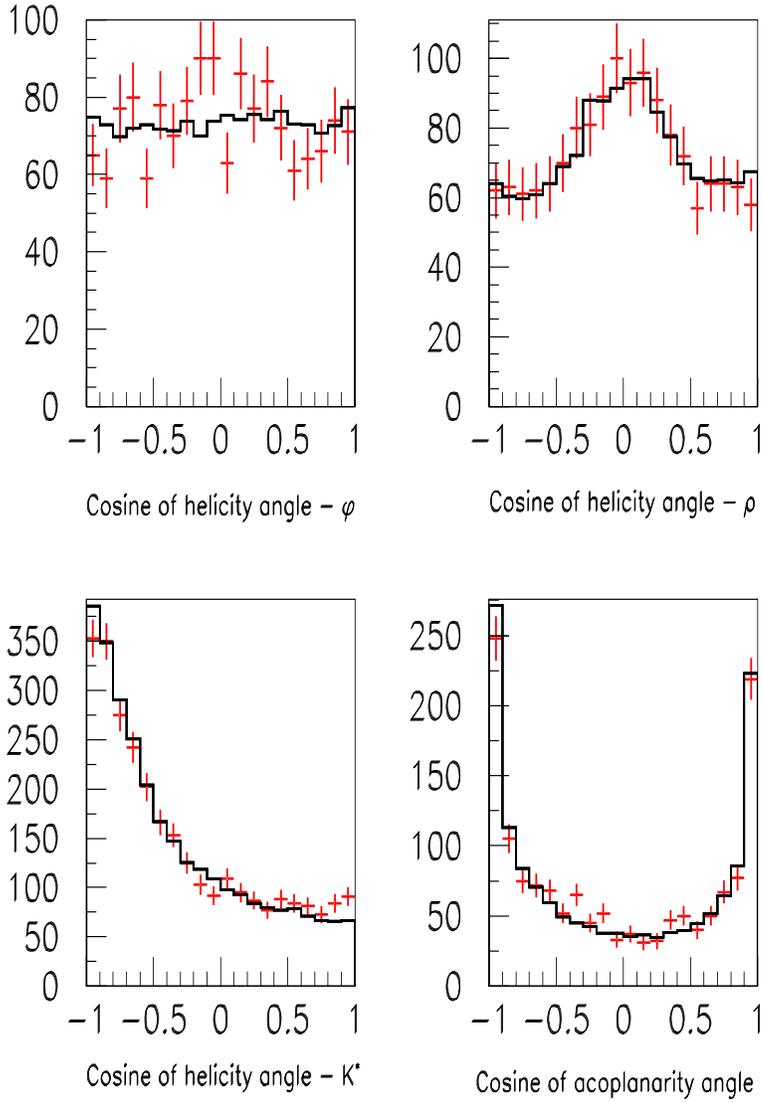} \vspace{0.5cm}
\caption{Distribution of cosine of the helicity and acoplanarity angles 
(see text for angle definitions)
The bump in the central part in the distribution of cosine of $\rho$ helicity
angle (top right plot)
is caused by the chain $D^0 \to K_1(1270)K$, $K_1(1270) \to \rho K$.}
\label{ang}
\end{figure}
\newpage

\end{document}